\documentclass[prl,reprint,showpacs,floatfix]{revtex4-1}

\usepackage{graphicx}
\usepackage{dcolumn}
\usepackage{bm}
\usepackage{color}
\usepackage{url}
\usepackage{amsmath}
\usepackage{amssymb}
\usepackage{mathrsfs}

\begin{document}

\title{Kinematical Test of Large Extra Dimension in Beta Decay Experiments}

\author{Victor S. Basto-Gonzalez$^1$}
\email{vsbasto@ifi.unicamp.br}
\author{Arman Esmaili$^1$}
\email{aesmaili@ifi.unicamp.br}
\author{Orlando L. G. Peres$^{1,2}$}
\email{orlando@ifi.unicamp.br}
\affiliation{$^1$Instituto de F\'isica Gleb Wataghin - UNICAMP, 13083-859, Campinas, SP, Brazil\ }
\affiliation{$^2$The Abdus Salam International Centre for Theoretical Physics, I-34100 Trieste, Italy}

\begin{abstract}
The forthcoming experiments on neutrino mass measurement using beta decay, open a new window
to explore the Large Extra Dimension model. The Kaluza-Klein tower of neutrinos in Large Extra Dimension
contribute to the Kurie function of beta decay that can be tested kinematically.
In addition to providing an alternative approach using just the kinematical properties,
we show that KATRIN can probe the compactification radius of extra dimensions down to 0.2 $\mu$m
which is better, at least by a factor of two, than the upper limits from neutrino oscillation experiments.
\end{abstract}

\maketitle

The ``hierarchy problem" is one of the long standing problems of the Standard Model (SM) of particles which triggered several beyond standard model theories (such as supersymmetry, technicolor, $\ldots$) as a way out. As a simple description, the hierarchy problem is the large disparity between the weak interaction scale $M_{\rm EW}\sim10^3$~GeV and the scale at which the gravity is strong, the so-called Planck scale $M_{\rm Pl}\sim10^{19}$~GeV. Within the SM particles, the anticipated Higgs particle suffers from this hierarchy through the radiative corrections to its mass which result in a Higgs mass of the order of Planck mass, contrary to the expectation from electroweak precision tests.

The theory of Large Extra Dimensions (LED) proposed in the seminal papers~\cite{ArkaniHamed:1998rs,ArkaniHamed:1998nn} as an elegant solution to the hierarchy problem. The basic assumption in LED is that there is just one fundamental scale, which is the weak scale $M_{\rm EW}$, and the large value of the observed $M_{\rm Pl}$ from gravity is a manifestation of the existence of extra dimensions such that $M_{\rm Pl}\sim M_{\rm EW}$ in the higher dimensional space but effectively is large in our 4-dimensional space. Assuming the existence of $n$ compactified extra dimensions, it can be shown that gravity would deviate from the inverse-square law at distances $\sim 10^{\frac{30}{n}-17} ({\rm TeV}/M_{\rm EW})^{1+\frac{2}{n}}\, {\rm cm}$~\cite{ArkaniHamed:1998rs}. Thus, to avoid the conflict with the confirmed inverse-square behavior of gravity at solar scale, the number of extra dimensions should be $n\geq 2$. However, here we assume an asymmetric space such that one out of the $n$ extra dimensions is compactified on a larger spatial scale $R_{\rm ED}$ and effectively we are facing a 5-dimensional space. 

The setup of LED model is as follows: the particles with charge under the SM gauge group including the charged leptons, active neutrinos, quarks, gauge bosons and Higgs particle live on a 4-dimensional brane embedded in the $(4+n)$-dimensional space. Several mechanism can justify the localization of these particles~\cite{ArkaniHamed:1998rs,Antoniadis:1998ig}, but to justify $M_{\rm Pl}\sim M_{\rm EW}$ (weakness of gravity) it is assumed that the singlets of the SM gauge group including the mediator of gravity (graviton) can propagate freely in the extra dimensions.

A bonus of the LED model is to give a natural explanation for the smallness of neutrino masses~\cite{ArkaniHamed:1998vp,Dienes:1998sb,Dvali:1999cn,Barbieri:2000mg}. As a concrete model, let us consider the LED extension of the SM augmented by three massless right-handed 5-dimensional neutrinos $\Psi^\alpha$ corresponding to the three active neutrino flavors. To be singlet under the SM gauge group, enable the $\Psi^\alpha$ fields to propagate in the 5-dimensional $(x^\mu,y)$ space, where $y$ denotes the large extra dimension ($\mu=0,1,2,3$). The general action describing the neutrino sector (assuming lepton number conservation by assigning lepton number -1 to $\Psi^\alpha$) is given by~\cite{ArkaniHamed:1998vp,Dienes:1998sb,Dvali:1999cn}
\begin{eqnarray}\label{action}
    S &=& \int {\rm d}^4x \,{\rm d}y \, i \overline{\Psi}^\alpha\Gamma_\text{A}\partial^\text{A}\Psi^\alpha + \int {\rm d}^4x\, \big\{ i \bar{\nu}_\text{L}^\alpha \gamma_\mu\partial^\mu\nu_\text{L}^\alpha   \nonumber \\
    &+& \; \kappa_{\alpha\beta} H\bar{\nu}_\text{L}^\alpha\psi^\beta_\text{R}(x,y=0) + \text{H.c.}\big\},
\end{eqnarray}
where $A=0,\ldots,4$; $H$ is the Higgs doublet, $\kappa$ is the Yukawa coupling matrix and the right-handed neutrino field is decomposed to $(\psi_{\rm L}^\alpha,\psi_{\rm R}^\alpha)$. After electroweak symmetry breaking, the above action results in the $3\times 3$ Dirac neutrino mass matrix $\sim \kappa v /\sqrt{V_n M_{\rm EW}^n}$, where $v$ is the VEV of Higgs doublet and $V_n$ is the volume of the extra $n$-dimensional space which naturally suppresses the mass matrix. Rotating the neutrino states to the mass eigenstates, gives the three eigenvalues of the $3\times 3$ mass matrix $(m_1^{\rm D},m_2^{\rm D},m_3^{\rm D})$.

From the 4-dimensional brane point of view, the right-handed neutrinos $\Psi^\alpha$ appear as a tower of Kaluza-Klein (KK) modes with increasing masses. Taking into account the quantized Dirac masse of these KK modes, the eigenvalues of the resulting infinite-dimensional mass matrix gives the masses of the neutrino mass eigenstates. The eigenvalues are given by $\lambda_i/R_{\rm ED}$ ($i=1,2,3$), where $\lambda_i$ is the root of the following transcendental equation~\cite{Dvali:1999cn,Barbieri:2000mg,Dienes:1998sb}
\begin{equation}\label{lambda}
\lambda_i -\pi R_{\rm ED}^2 \left(m_i^{\rm D}\right)^2 \cot (\pi \lambda_i)=0~.
\end{equation}
It is easy to check that Eq.~(\ref{lambda}) has infinite number of solutions $\lambda_i^{(n)}$, where $n=0,1,\ldots$; and $\lambda_i^{(n)}\in [n,n+1/2]$. It should be emphasized that $(m_1^{\rm D},m_2^{\rm D},m_3^{\rm D})$ are the parameters of the action in Eq.~(\ref{action}) and the masses of the neutrino states (including the three active SM neutrinos and the KK tower of sterile neutrinos) are given by $\lambda_i^{(n)}/R_{\rm ED}$ ($i=1,2,3$), where $n=0$ and $n=1,\ldots,$ corresponds respectively to the active and sterile neutrinos. However, the two solar and atmospheric mass-squared differences ($\Delta m_{\rm sol}^2,\Delta m_{\rm atm}^2$), from neutrino oscillation phenomenology, fixes two out of the three parameters $(m_1^{\rm D},m_2^{\rm D},m_3^{\rm D})$. The procedure is as follows: regarding $m_1^{\rm D}$ and $R_{\rm ED}$ as free parameters, it is possible calculate the value of $\lambda_1^{(0)}$ from Eq.~(\ref{lambda}) for fixed values of $m_1^{\rm D}$ and $R_{\rm ED}$. Then, the values of $\lambda_2^{(0)}$ and $\lambda_3^{(0)}$ is given by 
\begin{equation}\label{NH1}
\left(\lambda_2^{(0)}\right)^2=\left(\lambda_1^{(0)}\right)^2+R_{\rm ED}^2\Delta m_{\rm sol}^2~,
\end{equation}
and 
\begin{equation}\label{NH2}
\left(\lambda_3^{(0)}\right)^2=\left(\lambda_1^{(0)}\right)^2+R_{\rm ED}^2\Delta m_{\rm atm}^2~,
\end{equation}
for the normal hierarchy (NH) scheme on neutrino masses ($\lambda_3^{(0)}>\lambda_2^{(0)}>\lambda_1^{(0)}$). For the inverted hierarchy (IH) scheme ($\lambda_2^{(0)}>\lambda_1^{(0)}>\lambda_3^{(0)}$), regarding $m_3^{\rm D}$ and $R_{\rm ED}$ as free parameters, $\lambda_3^{(0)}$ can be calculated from Eq.~(\ref{lambda}), and  
\begin{equation}\label{IH}
\left(\lambda_2^{(0)}\right)^2\simeq\left(\lambda_1^{(0)}\right)^2=\left(\lambda_3^{(0)}\right)^2+R_{\rm ED}^2\Delta m_{\rm atm}^2~.
\end{equation}
Having the values of $\lambda_2^{(0)}$ and $\lambda_3^{(0)}$ for NH ($\lambda_1^{(0)}$ and $\lambda_2^{(0)}$ for IH), it is possible to read the value of $m_2^{\rm D}$ and $m_3^{\rm D}$ (or $m_1^{\rm D}$ and $m_2^{\rm D}$ for IH) from the Eq.~(\ref{lambda}). After that, the masses of sterile states in the KK towers can be calculated by finding the roots $\lambda_i^{(n\neq0)}$ of Eq.~(\ref{lambda}). Thus, the only free parameters in the mass matrix of the action Eq.~(\ref{action}) are the $(m_1^{\rm D},R_{\rm ED})$ for NH and $(m_3^{\rm D},R_{\rm ED})$ for IH. 

\begin{figure}[t]
\includegraphics[angle=-90,scale=0.35]{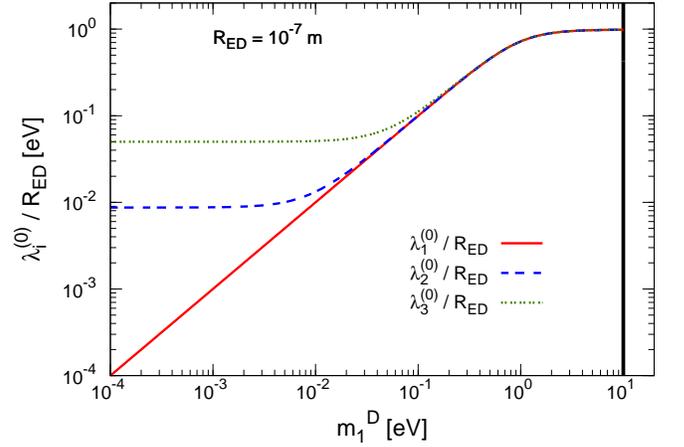}
\caption{\label{fig:lambda} 
The plot of $\lambda_i^{(0)}/R_{\rm ED}$ ($i=1,2,3$) as a function of $m_1^{\rm D}$ assuming NH scheme and $R_{\rm ED}=10^{-7}\,{\rm m}$.
}
\end{figure}

However, in deriving the values of $\lambda_2^{(0)}$ and $\lambda_3^{(0)}$ from Eqs.~(\ref{NH1},\ref{NH2}), or $\lambda_1^{(0)}$ and $\lambda_2^{(0)}$ from Eq.~(\ref{IH}) in the case of IH, we are constraint by the fact that for any values of $(\Delta m_{\rm sol}^2,\Delta m_{\rm atm}^2)$ always we should have $\lambda_i^{(0)}\leq 0.5$ (see the comment after Eq.~(\ref{lambda})). This inequality defines the allowed physical region in the $(m_1^{\rm D},R_{\rm ED})$ or $(m_3^{\rm D},R_{\rm ED})$ parameter space which a solution exist to the observed solar and atmospheric oscillation scales within the framework of LED model. Fig.~\ref{fig:lambda} shows the behavior of $\lambda_i^{(0)}/R_{\rm ED}$ as a function of $m_1^{\rm D}$ assuming NH scheme for $(\Delta m_{\rm sol}^2,\Delta m_{\rm atm}^2)=(7.6\times10^{-5},2.4\times10^{-3})~{\rm eV}^2$~\cite{Schwetz:2011qt} and $R_{\rm ED}=10^{-7}$~m. As can be seen, for $m_1^{\rm D}\ll 1/R_{\rm ED}$ we have $\lambda_1^{(0)}/R_{\rm ED}\simeq m_1^{\rm D}$ which comes from the fact that in this case the KK tower sterile neutrinos have a very small contribution to the neutrino mass matrix. But, however, for $m_1^{\rm D}\simeq 1/R_{\rm ED}$ the mass $\lambda_1^{(0)}/R_{\rm ED}$ saturate to the value $0.5/R_{\rm ED}$~. The black vertical thick line in Fig.~\ref{fig:lambda} shows the discussed allowed region coming from the condition $\lambda_3^{(0)}\leq 0.5$~.

\begin{figure}[b]
\includegraphics[angle=-90,scale=0.35]{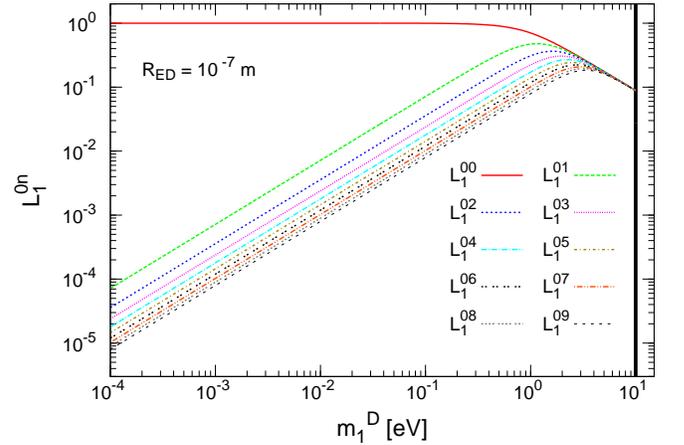}
\caption{\label{fig:lplot} 
The plot of $L_1^{0n}$ as a function of $m_1^{\rm D}$ assuming $R_{\rm ED}~=~10^{-7}\,{\rm m}$. The curves from top to bottom correspond to $n=0,\ldots,9$. 
}
\end{figure}

The contribution of the KK tower sterile neutrinos to the neutrino mass matrix leads to the mixing between the active and sterile neutrinos. The active flavor neutrino states can be written as~\cite{Dvali:1999cn,Barbieri:2000mg,Dienes:1998sb}
\begin{equation} \label{eq:mix}
\nu_{\rm L}^\alpha = \sum_{i = 1}^3U^{\alpha i *} \sum_{n = 0}^\infty L_i^{0n} \nu_\text{L}^{i(n)}~,
\end{equation}
where $\nu_\text{L}^{i(n)}$ are the mass eigenstates composed of the $n^{\rm th}$ KK mode of $\psi_{\rm L}$ ($n=0$ corresponds to the SM active mass eigenstates) and the $3\times 3$ matrix $U$ is the Pontecorvo-Maki-Nakagawa-Sakata mixing matrix of neutrinos~\cite{Maki:1962mu}. The parameters $L_i^{0n}$ come from the diagonalization of the infinite-dimensional mass matrix~\cite{Dienes:1998sb,Dvali:1999cn,Barbieri:2000mg} and is given by
\begin{equation}\label{eq:l}
(L_i^{0n})^2 = {2\over{1 + \pi^2\left(R_{\rm ED}m_i^{\rm D}\right)^2 + \lambda_i^{(n)2}/\left(R_{\rm ED}m_i^{\rm D}\right)^2}}~.
\end{equation}
The values of the $L_i^{0n}$ parameters also is just a function of the two free parameters $(m_1^{\rm D},R_{\rm ED})$ for NH and $(m_3^{\rm D},R_{\rm ED})$ for IH. To clarify the behavior of the parameters in Eq.~(\ref{eq:l}) we plotted in Fig.~\ref{fig:lplot} the $L_1^{0n}$ for $n=0,\ldots,9$; as a function of $m_1^{\rm D}$ and assuming $R_{\rm ED}~=~10^{-7}\,{\rm m}$. As can be seen, for $m_1^{\rm D}\ll 1/R_{\rm ED}$, the values of $L_1^{0n\neq0}$ is very small which means that the contribution of the sterile mass eigenstates to the decomposition of active flavor states in Eq.~(\ref{eq:mix}) is almost negligible. But, for $m_1^{\rm D}\simeq 1/R_{\rm ED}$, KK tower states with higher masses play role in the active states decomposition.

The decomposition of the flavor neutrino states in Eq.~(\ref{eq:mix}) would impact the phenomenology of the neutrino sector of SM in a substantial way. From flavor oscillation point of view, considering the decomposition in Eq.~(\ref{eq:mix}), the probability of $\nu_{\rm L}^\alpha\to \nu_{\rm L}^\beta$ oscillation for neutrinos with energy $E_\nu$ at the baseline $L$ is given by~\cite{Dvali:1999cn,Barbieri:2000mg} 
\begin{equation}\label{eq:prob}
P_{\alpha\beta} =\Bigg\vert \sum_{i = 1}^{3}U^{\alpha i}U^{\beta i\ast}\sum_{n = 0}^\infty (L_i^{0n})^2 \exp{\left[i\frac{\lambda_i^{(n)2}L}{2E_\nu R_{\text{ED}}^2}\right]\Bigg\vert^2}.
\end{equation} 
Plenty of works is devoted to use of the above formula in fitting the data of oscillation experiments~\cite{Barbieri:2000mg,Davoudiasl:2002fq,Mohapatra:1999zd}, and recently~\cite{Machado:2011jt} . The updated result including CHOOZ, MINOS and KamLAND experiments gives $R_{\rm ED}\lesssim 6\times~10^{-7}\,{\rm m}$ at 90\% C.~L.~\cite{Machado:2011jt}.

From the kinematical point of view, the decomposition in Eq.~(\ref{eq:mix}) would impact the neutrino mass measurement experiments. Specifically, the KK tower sterile neutrino states contribute to the effective electron anti-neutrino mass which can be probed at the beta decay experiments. The forthcoming KArlsruhe TRItium Neutrino (KATRIN) experiment~\cite{katrin}, probes the kinematical effect of the $\bar{\nu}_e$ produced in the tritium beta decay ${}^3$H$\to{}^3$He${}^+ + e^- + \bar{\nu}_e$ by the measurement of electron energy spectrum near the endpoint of the reaction $Q=18571.8\pm1.2 \; {\rm eV}$~\cite{qvalue}. In the absence of neutrino mass, the Kurie function of beta decay is a linear function of the electron's kinetic energy $T_e$, that is $K(T_e)=Q-T_e$. A nonzero value of the neutrino mass leads to distortion in the Kurie function, such that the endpoint of the electron's energy spectrum displaces by the lightest neutrino mass value and heavier masses generate kinks in the spectrum. Assuming the decomposition of Eq.~(\ref{eq:mix}) for electron anti-neutrino, the Kurie function takes the following form 
\begin{eqnarray}\label{eq:kurie}
K(T_e,m_0,R_{\rm ED}) =& & \\
& \!\!\!\!\!\!\!\!\!\!\!\!\!\!\!\!\!\!\!\!\!\!\!\!\!\!\!\!\!\!\!\!\!\!\!\!\!\!\!\!\!\!\! \displaystyle \sum_k p_k\mathcal{E}_k \sum_{i=1}^{3} |U^{ei}|^2 \sum_{n=0}^{\infty} (L_{i}^{0n})^2 \sqrt{\mathcal{E}_k^2-\left(\frac{\lambda_i^{(n)}}{R_{\rm ED}}\right)^2}~, & \nonumber
\end{eqnarray} 
where $\mathcal{E}_k=Q-W_k-T_e$; and hereafter $m_0\equiv m_1^{\rm D}$ for NH and $m_0\equiv m_3^{\rm D}$ for IH. The dependence on $m_0$ comes from the $\lambda_i^{(n)}$ ({\it see} Eq.~(\ref{lambda})). The $W_k$ and $p_k$ are respectively the excitation energy and transition probability of the $k^{\rm th}$ rotational and vibrational excited state of the daughter nucleus in the tritium beta decay~\cite{excite}. Also, inserting the Heaviside step function $\Theta(\mathcal{E}_k-\lambda_i^{(n)}/R_{\rm ED})$ in Eq.~(\ref{eq:kurie}) guarantees the conservation of energy. 

The spectrometer of the KATRIN experiment is accumulative, which means that it can collect the electrons with kinetic energy larger than the electrostatic barrier $qU$ which can be tuned. The rate of the electrons passing the potential barrier is given by 
\begin{eqnarray}
&\!\!\!\!\!\!\!\!\!\!\!\!\!\!\!\!\!\!\!\!\!\!\!\!\!\!\!\!\!\!\!\!\!\!\!\!\!\!\!\!\!\!\!\!\!\!\!\!\!\!\!\!\!\!\!\!\!\!\!\!\!\!\!\!\!\!\!\!\!\!\!\!\!\!\! S(qU,m_0,R_{\rm ED})=N_b+\\ 
&\qquad\!\!\!\!\!  \displaystyle N_s \int_0^\infty
  F(Z,T_e) E_e p_e K(T_e,m_0,R_{\rm ED}) R^\prime (T_e,qU)\, {\rm d}T_e \, , \nonumber
\label{spectrum}
\end{eqnarray}
where $E_e$ and $p_e$ are respectively the energy and momentum of the electron. The $F(Z,T_e)$ is the Fermi function which takes into account the electromagnetic interaction of the emitted electron in beta decay with the daughter nucleus ($Z=2$)~\cite{ffunc}. The function $R^\prime$ shows the transmission function of the KATRIN experiment including the resolution of the spectrometer and the energy loss processes in the propagation of electrons from the source to collector. Finally, the $N_s$ is the normalization factor calibrating the rate of electron emission which for KATRIN is $1.47\times
10^{-13}\,{\rm s}^{-1}\,{\rm eV}^{-5}$; and $N_b$ is the background rate which is 10~mHz~\cite{katrin}. The details of the KATRIN experiment and functional form of $R^\prime$ can found in~\cite{Esmaili:2012vg}.

\begin{figure}[b]
\includegraphics[angle=-90,scale=0.35]{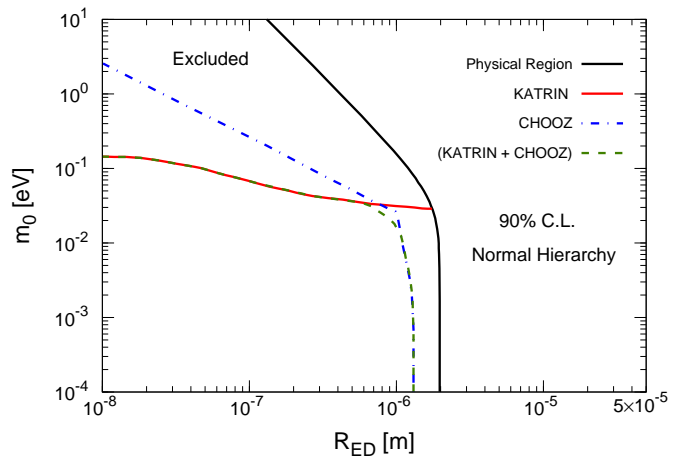}
\caption{\label{fig:senNH} 
The 90\% C. L. sensitivity in $(m_0,R_{\rm ED})$ parameter space for KATRIN experiment assuming NH for the neutrino mass scheme. Also we show the 90\% C. L. limit form CHOOZ experiment and the combined analysis of KATRIN+CHOOZ. 
}
\end{figure}
We calculated the sensitivity of KATRIN experiment, after three years of data-taking, using the proposed optimized running time for a barrier potential $qU\in[Q-20,Q+5]~{\rm eV}$~({\it see}~\cite{katrin}). Fig.~\ref{fig:senNH} shows the 90\%~C.~L. sensitivity to the $(m_0,R_{\rm ED})$ parameters assuming NH for the neutrino mass scheme. The black solid line shows the boundary of physical region of parameters which comes from the requirement that $\lambda_i^{(0)}\leq 0.5$~; {\it i.e.}, the values of $(m_0,R_{\rm ED})$ parameters in the right-hand side of the black curve is not compatible with $(\Delta m_{\rm sol}^2,\Delta m_{\rm atm}^2)=(7.6\times10^{-5},2.4\times10^{-3})~{\rm eV}^2$.  The red solid curve shows the 90\%~C.~L. sensitivity of KATRIN (for NH). To compare with the limits from oscillation experiments, we calculated the limit from CHOOZ experiment (blue dashed-dotted curve), which is the strongest limit within oscillation experiments. 
For the CHOOZ experiment~\cite{Apollonio:2002gd}, which is a reactor $\bar{\nu}_e$ disappearance experiment at the base-line $\sim 1\,{\rm km}$, we fit the obtained data using Eq.~(\ref{eq:prob}) for $\alpha=\beta=e$. As can be seen, the KATRIN sensitivity is about one order of magnitude stronger in $R_{\rm ED}\lesssim 10^{-7}$~m region; but weaker for $R_{\rm ED}\gtrsim 10^{-6}$~m. To take advantage of the oscillation experiments, we did the combined analysis of KATRIN+CHOOZ which is shown by the green dashed line.

\begin{figure}[t]
\includegraphics[angle=-90,scale=0.35]{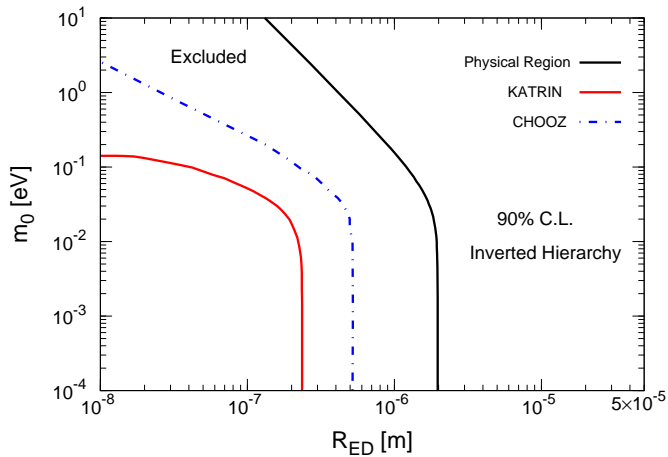}
\caption{\label{fig:senIH} 
The same as Fig.~\ref{fig:senNH} but for IH for neutrino mass scheme. Also, in this case the combined KATRIN+CHOOZ is the same as KATRIN alone. 
}
\end{figure}

In Fig.~\ref{fig:senIH} we show the 90\%~C.~L. sensitivity of KATRIN for the IH case. As can be seen, for all the values of $m_0$ and $R_{\rm ED}$, the KATRIN sensitivity is stronger than the CHOOZ limit which is the strongest within the oscillation experiments. Also, because of the stronger sensitivity in KATRIN, the combined analysis of CHOOZ+KATRIN gives the same result as KATRIN alone.   

The KATRIN sensitivity is much stronger than the limits coming from probing deviations of gravity law from inverse-square at sub-millimeter scales, which is $R_{\rm ED}\leq 1.6\times10^{-4}$~m~\cite{Hoyle:2004cw}. On the other hand, strong limits on the size of extra dimensions exist from the astrophysical and cosmological considerations. The astrophysical limits stem from cooling time of supernovae~\cite{Cullen:1999hc} and diffuse gamma rays from the decay of KK gravitons produced at the supernovae~\cite{Hannestad:2003yd}. For the number of extra dimensions $n=2$ ($n=3$), the upper limit from cooling of supernovae is $R_{\rm ED}\leq 0.66\times10^{-6}$~m ($R_{\rm ED}\leq 0.8\times10^{-9}$~m) and the limit from KK gravitons decay to photons is $R_{\rm ED}\leq 3.61\times10^{-7}$~m ($R_{\rm ED}\leq 3.95\times10^{-10}$~m). The production of KK modes in the early universe also can significantly affect the cosmological evolution~\cite{Hall:1999mk}. The limits from this cosmological consideration is $R_{\rm ED}\leq 2.2\times10^{-8}$~m ($R_{\rm ED}\leq 2.5\times10^{-11}$~m) for $n=2$ ($n=3$). However, it should be noticed that all of these limits from astrphysics and cosmology suffer from model dependency. For example, it can be shown that compactification of extra dimensions on a hyperbolic manifold relaxes all the above mentioned limits completely~\cite{Kaloper:2000jb}. Specifically, for the model we are considering in this paper, with an asymmetrical compactification of extra dimensions, the limits from astrophysics and cosmology do not apply ({\it see} also~\cite{Giudice:2004mg}) The robustness of the limit from forthcoming KATRIN experiment is based on the fact that limit comes just from kinematical considerations and do not suffer from theoretical uncertainties. In this sense, the KATRIN experiment triggers an alternative approach to the test of large extra dimension model.

In summary, we have shown that the forthcoming neutrino mass measurement experiments, using the beta decay (such as KATRIN), can test the LED model extended with singlet 5-dimensional neutrino fields that can propagate in the bulk. This extension was motivated as an elegant way to justify the smallness of neutrino masses. Besides the fact that this kinematical test at KATRIN is a new window to probe LED, we have shown that this test can give stronger limit on the compactification scale $R_{\rm ED}$. For the case of inverted hierarchy in the neutrino mass scheme, KATRIN can obtain the upper limit $R_{\rm ED}\leq 2.3\times10^{-7}$~m (90\%~C.~L.) after three years of data-taking (for $m_0\to0$); which is a factor of two better than the limit from oscillation experiments. Also, for the normal hierarchy scheme, KATRIN can exclude the regions of the $(m_0,R_{\rm ED})$ parameter space which is inaccessible to oscillation experiments.

\begin{acknowledgments}
{\small A.~E. and O.~L.~G.~P.  thank support from FAPESP. O.~L.~G.~P. thanks support from CAPES/Fulbright. V.~S.~B.~G. thanks support from CNPq. 
The authors acknowledge the use of CENAPAD-SP and CCJDR computing facilities.
}
\end{acknowledgments}


\begin{thebibliography}{30}

\bibitem{ArkaniHamed:1998rs} 
  N.~Arkani-Hamed, S.~Dimopoulos and G.~R.~Dvali,
  Phys.\ Lett.\ B {\bf 429}, 263 (1998)
  [hep-ph/9803315].
  
\bibitem{ArkaniHamed:1998nn} 
  N.~Arkani-Hamed, S.~Dimopoulos and G.~R.~Dvali,
  Phys.\ Rev.\ D {\bf 59}, 086004 (1999)
  [hep-ph/9807344].
  
\bibitem{Antoniadis:1998ig} 
  I.~Antoniadis, N.~Arkani-Hamed, S.~Dimopoulos and G.~R.~Dvali,
  Phys.\ Lett.\ B {\bf 436}, 257 (1998)
  [hep-ph/9804398].
  
\bibitem{ArkaniHamed:1998vp} 
  N.~Arkani-Hamed, S.~Dimopoulos, G.~R.~Dvali and J.~March-Russell,
  Phys.\ Rev.\ D {\bf 65}, 024032 (2002)
  [hep-ph/9811448].
  
\bibitem{Dienes:1998sb} 
  K.~R.~Dienes, E.~Dudas and T.~Gherghetta,
  Nucl.\ Phys.\ B {\bf 557}, 25 (1999)
  [hep-ph/9811428].
  
\bibitem{Dvali:1999cn} 
  G.~R.~Dvali and A.~Y.~.Smirnov,
  Nucl.\ Phys.\ B {\bf 563}, 63 (1999)
  [hep-ph/9904211].
  
\bibitem{Barbieri:2000mg} 
  R.~Barbieri, P.~Creminelli and A.~Strumia,
  Nucl.\ Phys.\ B {\bf 585}, 28 (2000)
  [hep-ph/0002199].
  
\bibitem{Schwetz:2011qt} 
  T.~Schwetz, M.~Tortola and J.~W.~F.~Valle,
  New J.\ Phys.\  {\bf 13}, 063004 (2011)
  [arXiv:1103.0734 [hep-ph]].
  
\bibitem{Maki:1962mu} 
  Z.~Maki, M.~Nakagawa and S.~Sakata,
  Prog.\ Theor.\ Phys.\  {\bf 28}, 870 (1962);
  B.~Pontecorvo,
  Sov.\ Phys.\ JETP {\bf 6}, 429 (1957)
  [Zh.\ Eksp.\ Teor.\ Fiz.\  {\bf 33}, 549 (1957)].
  
\bibitem{Davoudiasl:2002fq}
  H.~Davoudiasl, P.~Langacker and M.~Perelstein,
  Phys.\ Rev.\ D {\bf 65}, 105015 (2002)
  [hep-ph/0201128].
  
\bibitem{Mohapatra:1999zd} 
  R.~N.~Mohapatra, S.~Nandi and A.~Perez-Lorenzana,
  Phys.\ Lett.\ B {\bf 466}, 115 (1999)
  [hep-ph/9907520];
  R.~N.~Mohapatra and A.~Perez-Lorenzana,
  Nucl.\ Phys.\ B {\bf 576}, 466 (2000)
  [hep-ph/9910474];
  R.~N.~Mohapatra and A.~Perez-Lorenzana,
  Nucl.\ Phys.\ B {\bf 593}, 451 (2001)
  [hep-ph/0006278].
  
\bibitem{Machado:2011jt}
  P.~A.~N.~Machado, H.~Nunokawa and R.~Zukanovich Funchal,
  Phys.\ Rev.\ D {\bf 84}, 013003 (2011)
  [arXiv:1101.0003 [hep-ph]].
  
\bibitem{katrin} 
  J.~Angrik {\it et al.}  [KATRIN Collaboration],
  FZKA-7090.
  
\bibitem{qvalue}
R.~Schuch  {\it et al.} 
Hyperfine Interactions {\bf 173}, 73 (2006).  

\bibitem{excite}
  A.~Saenz, S.~Jonsell and P.~Froelich,    
  Phys.\ Rev.\ Lett.\ {\bf 84}, 242 (2000).
  
 \bibitem{ffunc}
  V.~N.~Aseev {\it et al.} 
Eur.\ Phys.\ J.\ D {\bf 10} (1) 39-52 (2000).

\bibitem{Esmaili:2012vg} 
  A.~Esmaili and O.~L.~G.~Peres,
  Phys.\ Rev.\ D {\bf 85}, 117301 (2012)
  [arXiv:1203.2632 [hep-ph]].
  
\bibitem{Apollonio:2002gd}
  M.~Apollonio {\it et al.}  [CHOOZ Collaboration],
  Eur.\ Phys.\ J.\ C {\bf 27}, 331 (2003)
  [hep-ex/0301017].
  
    
\bibitem{Hoyle:2004cw} 
  C.~D.~Hoyle {\it et al.},
  Phys.\ Rev.\ D {\bf 70}, 042004 (2004)
  [hep-ph/0405262].

\bibitem{Cullen:1999hc} 
  S.~Cullen and M.~Perelstein,
  Phys.\ Rev.\ Lett.\  {\bf 83}, 268 (1999)
  [hep-ph/9903422];
  V.~D.~Barger, T.~Han, C.~Kao and R.~-J.~Zhang,
  Phys.\ Lett.\ B {\bf 461}, 34 (1999)
  [hep-ph/9905474];
  C.~Hanhart, D.~R.~Phillips, S.~Reddy and M.~J.~Savage,
  Nucl.\ Phys.\ B {\bf 595}, 335 (2001)
  [nucl-th/0007016];
  C.~Hanhart, J.~A.~Pons, D.~R.~Phillips and S.~Reddy,
  Phys.\ Lett.\ B {\bf 509}, 1 (2001)
  [astro-ph/0102063].
  

\bibitem{Hannestad:2003yd}
  S.~Hannestad and G.~Raffelt,
  Phys.\ Rev.\ Lett.\  {\bf 87}, 051301 (2001)
  [hep-ph/0103201];
  S.~Hannestad and G.~G.~Raffelt,
  Phys.\ Rev.\ Lett.\  {\bf 88}, 071301 (2002)
  [hep-ph/0110067];
  S.~Hannestad and G.~G.~Raffelt,
  Phys.\ Rev.\ D {\bf 67}, 125008 (2003)
  [Erratum-ibid.\ D {\bf 69}, 029901 (2004)]
  [hep-ph/0304029].
  
\bibitem{Hall:1999mk} 
  L.~J.~Hall and D.~Tucker-Smith,
  Phys.\ Rev.\ D {\bf 60}, 085008 (1999)
  [hep-ph/9904267];
  M.~Fairbairn,
  Phys.\ Lett.\ B {\bf 508}, 335 (2001)
  [hep-ph/0101131];
  S.~Hannestad,
  Phys.\ Rev.\ D {\bf 64}, 023515 (2001)
  [hep-ph/0102290].
  
  
\bibitem{Kaloper:2000jb} 
  N.~Kaloper, J.~March-Russell, G.~D.~Starkman and M.~Trodden,
  Phys.\ Rev.\ Lett.\  {\bf 85}, 928 (2000)
  [hep-ph/0002001];
  G.~D.~Starkman, D.~Stojkovic and M.~Trodden,
  Phys.\ Rev.\ D {\bf 63}, 103511 (2001)
  [hep-th/0012226].
  
\bibitem{Giudice:2004mg} 
  G.~F.~Giudice, T.~Plehn and A.~Strumia,
  Nucl.\ Phys.\ B {\bf 706}, 455 (2005)
  [hep-ph/0408320].



\end{thebibliography}
\end{document}